\documentclass[twocolumn]{jpsj3}
\usepackage{color}

\title{
Structural Domain Switching by Magnetic Fields in RAl$_3$C$_3$ \\(R=Rare Earth)
}

\author{T. Matsumura$^{1,2}$, Y. Hayashi$^{1}$, S. Takai$^{1}$, T. Otsubo$^{1}$, 
S. Matsuda$^{3}$, and A. Ochiai$^{3}$}

\inst{$^{1}$Department of Quantum Matter, AdSM, Hiroshima University, Higashi-Hiroshima, 739-8530, Japan \\
$^{2}$Institute for Advanced Materials Research, Hiroshima University, Higashi-Hiroshima, 739-8530, Japan\\
$^{3}$Department of Physics, Tohoku University, Sendai 980-8578, Japan\\
}

\abst{
Hexagonal-to-orthorhombic structural transitions in RAl$_3$C$_3$ have been studied by X-ray diffraction for R=Yb, Tm, Er, Dy, and Lu. 
We report how orthorhombic domains are controlled by adjusting external magnetic fields. 
It is shown that orthorhombic domains in RAl$_3$C$_3$ can be aligned by the field when the R ion is magnetic. 
It is also shown that a single-domain state can be switched to another single-domain state by the field even in paramagnetic states with small induced moments. 
For R=Lu, however, it was not possible to change the multidomain state by a field of up to 70 kOe. 
We discuss that domain selection and switching are caused by magnetic anisotropies in the orthorhombic phase of RAl$_3$C$_3$. We also propose an idea to explain the field-induced antiferromagnetism revealed by Khalyavin \textit{et al.} [Phys. Rev. B \textbf{87}, 220406 (2013)]. 
 }

\begin{document}
\setlength{\textwidth}{504pt}
\setlength{\columnsep}{14pt}
\hoffset-23.5pt
\setlength{\textheight}{680pt}

\maketitle
\section{Introduction}
The formation of quantum spin singlet states in magnetic materials has attracted much attention. 
Although there are many quantum spin systems in $d$-electron compounds, it is very rare in $f$-electron compounds. 
This is because the total angular momentum $J$, as a result of a strong spin-orbit coupling, is much larger than 1/2. 
However, if the $J$-multiplet is split by a crystalline electric field (CEF) to form a Kramers doublet ground state, it can effectively be treated as a $S=1/2$ system. 
One typical example is Yb$_4$As$_3$, where a one-dimensional $S=1/2$ Heisenberg antiferromagnetic chain is realized in the charge-ordered state.\cite{Kohgi97,Kohgi01,Shiba00} 
Another candidate for a quantum spin singlet state in $f$-electron systems is YbAl$_3$C$_3$, which has recently been proposed to form a spin-dimer singlet.\cite{Ochiai07} 

RAl$_3$C$_3$  (R=rare earth) compounds have a hexagonal ScAl$_3$C$_3$-type structure with the space group $P6_3/mmc$ (No. 194).\cite{Gesing92} 
Rare-earth ions on the $c$-plane form a two-dimensional triangular lattice, which is well isolated from the next rare-earth layers by Al and C atoms in between. 
Interest in the RAl$_3$C$_3$ system arose from the discovery of a phase transition at $T_{\text{s}}=80$ K in YbAl$_3$C$_3$.\cite{Kosaka05} 
Although the transition was initially interpreted as a quadrupole ordering, it was later shown by X-ray diffraction to be a hexagonal-to-orthorhombic structural transition.\cite{Matsumura08} 
The most interesting feature of YbAl$_3$C$_3$ is its low temperature properties, which can approximately be interpreted using an isolated spin dimer model with the spin singlet ground state.\cite{Ochiai07,Hara12}
The triplet excited state at $\sim$15 K, which is expected from magnetic susceptibility and specific heat, has also been shown by inelastic neutron scattering.\cite{Kato08}
Recent interest in YbAl$_3$C$_3$ lies in detecting a field-induced ordered phase characteristic to a quantum spin system.\cite{Hara12,Khalyavin13}

The structural phase transition is a common feature in the RAl$_3$C$_3$ system. 
The physical properties of other RAl$_3$C$_3$ compounds also exhibit an anomaly at high temperatures corresponding to $T_{\text{s}}$ in YbAl$_3$C$_3$: 
110 K in LuAl$_3$C$_3$, 103 K in TmAl$_3$C$_3$, 120 K in ErAl$_3$C$_3$, and 50 K in DyAl$_3$C$_3$.\cite{Ochiai07,Ochiai10} 
Magnetic ordering occurs at 22 K in DyAl$_3$C$_3$ and at 13 K in ErAl$_3$C$_3$, whereas TmAl$_3$C$_3$ exhibits no magnetic ordering, suggesting a singlet CEF ground state. 

\begin{figure}[t]
\begin{center}
\includegraphics[width=8.5cm]{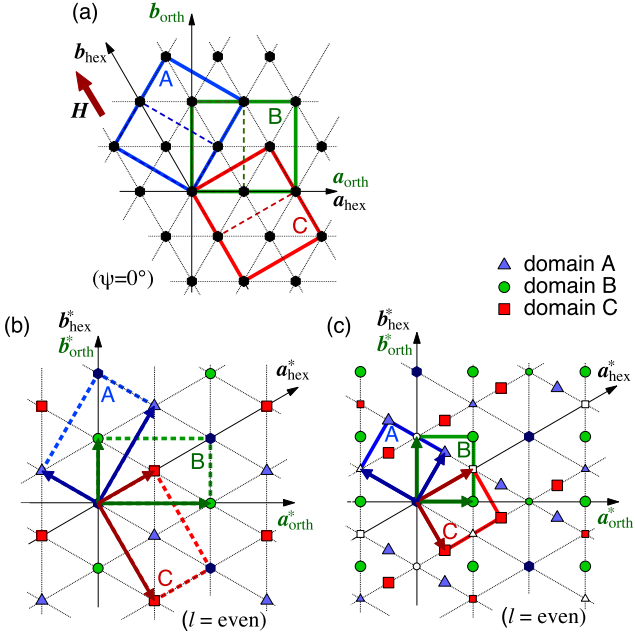}
\end{center}
\caption{(Color online) (a) Real-space unit cells of RAl$_3$C$_3$ in the $c$-plane. 
Closed hexagons represent lattice points. Dashed rectangles represent three domains of orthorhombic unit cells after the structural transitions in LuAl$_3$C$_3$, TmAl$_3$C$_3$, and ErAl$_3$C$_3$, and in DyAl$_3$C$_3$ in the intermediate phase. 
Solid rectangles represent those of YbAl$_3$C$_3$ and DyAl$_3$C$_3$ in the low-temperature phase. The three domains are denoted A, B, and C. 
(b) $hkl$ plane of the reciprocal space with $l$=even. Filled hexagons represent the Bragg spots of the fundamental hexagonal lattice. Filled triangles, circles, and squares correspond to those of domains A, B, and C, respectively, of the orthorhombic lattice corresponding to the dashed rectangles in (a). 
(c) Same as (b), but corresponding to the solid rectangles in (a). Open marks represent forbidden reflection points. }
\label{fig:Fig1}
\end{figure}

Another important point related to the structural transition is the domain selection by magnetic fields, which has been pointed out in YbAl$_3$C$_3$.\cite{Hara12}
In the low-temperature orthorhombic phase, there arise three structural domains as shown by the solid rectangles in Fig.~\ref{fig:Fig1}(a).\cite{Matsumura08}  
A history-dependent magnetization is observed in YbAl$_3$C$_3$ in this phase. This can be understood by considering that the initial multidomain state changes to a single-domain state by applying a magnetic field. 
It is therefore necessary to know the domain state to analyze the behavior in magnetic fields.  
The purpose of this study is to clarify the process of domain selection by magnetic fields in RAl$_3$C$_3$.

\section{Experiment}
Single-crystalline samples were prepared using sealed tungsten crucibles.\cite{Ochiai10} 
X-ray diffraction experiments were performed at the Photon Factory, High Energy Accelerator Research Organization (KEK), Japan. 
Hexagonal-slice-shaped single crystals with $c$-plane surfaces of $\sim 0.2\times 0.2$ mm$^2$ were used for the measurements. 
The search for superlattice Bragg peaks in the low-temperature phase was performed using an imaging-plate system at BL-8B. 
The positions of the Bragg peaks on the $hkl$ reciprocal lattice plane with $l$=even, which was clarified in this experiment, are summarized in Figs.~\ref{fig:Fig1}(b) and \ref{fig:Fig1}(c). 
Then, some selected Bragg peaks were investigated in magnetic fields using a vertical-field 8 Tesla superconducting magnet installed at BL-3A. Integrated intensities were measured by performing rocking scans. 
We used an X-ray energy of 18 keV at BL-8B and 14 keV at BL-3A. 
In the experiment at BL-3A, the samples were attached to a rotator, which was inserted in the cryomagnet. 
By rotating the sample around the $c$-axis, we can change the field direction in the $c$-plane as shown in Fig.~\ref{fig:Fig1}(a); 
the sample rotates counter clock wise when the angle $\psi$ increases. 

\section{Results} 
\subsection{YbAl$_3$C$_3$}
In the low-temperature orthorhombic phase, three structural domains arise, as shown by the solid rectangles in Fig.~\ref{fig:Fig1}(a), which represent the real-space unit cells in the $c$-plane.\cite{Matsumura08} 
The $hkl$ Bragg spots in the reciprocal space corresponding to these domains are summarized in Fig.~\ref{fig:Fig1}(c) by the filled marks for $l$=even. 

\begin{figure}[t]
\begin{center}
\includegraphics[width=8cm]{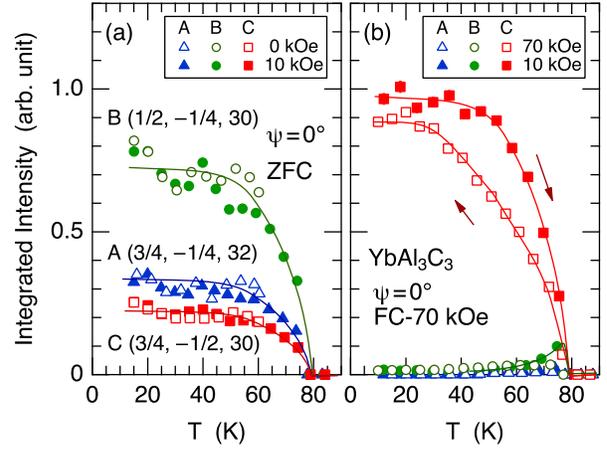}
\end{center}
\caption{(Color online) Temperature dependences of the superlattice peak intensities corresponding to the three orthorhombic domains of YbAl$_3$C$_3$ after (a) zero field cooling and (b) field cooling at 70 kOe. The field is applied along the hexagonal $b$-axis. 
}
\label{fig:Fig2}
\end{figure}

Figure~\ref{fig:Fig2} shows the temperature dependences of the superlattice peak intensities corresponding to the three orthorhombic domains of YbAl$_3$C$_3$. We measured the strong peaks with $l$=even. 
First, after zero field cooling (ZFC), the intensities were measured with increasing temperature. 
As shown in Fig.~\ref{fig:Fig2}(a) by the open marks, all the intensities for the three domains are observed. 
Second, after ZFC, a magnetic field of 10 kOe was applied at $\psi=0^{\circ}$ ($\mib{H}\parallel \mib{b}_{\text{hex}}$), and the intensities were measured with increasing temperature. 
As shown by the filled marks in Fig.~\ref{fig:Fig2}(a), the intensities are not affected by the field. 
This shows that the multidomain state is preserved in the field of 10 kOe after ZFC. 

By contrast, as shown in Fig.~\ref{fig:Fig2}(b), field cooling (FC) at 70 kOe ($\parallel\mib{b}_{\text{hex}}$) from the high-temperature hexagonal phase to the low-temperature orthorhombic phase selects the C-domain only. The intensities corresponding to the A- and B-domains are much smaller than those of the C-domain. 
After FC at 70 kOe, the field was decreased to 10 kOe at 10 K. 
In this case, the intensities of the A- and B-domains do not recover, indicating that the single-domain state is preserved after the field was reduced to 10 kOe. 
Then, the intensities were measured at 10 kOe with increasing temperature. As shown by the filled marks in Fig.~\ref{fig:Fig2}(b), the single-domain state is almost preserved up to $T_{\text{s}}=80$ K. 
This is the reason for the history-dependent magnetic properties of YbAl$_3$C$_3$ reported in Ref.~\citen{Hara12}. 
A slight increase in the B-domain intensity observed just below $T_{\text{s}}$ shows that domain repopulation occurs at high temperatures close to $T_{\text{s}}$. 

\begin{figure}[t]
\begin{center}
\includegraphics[width=8cm]{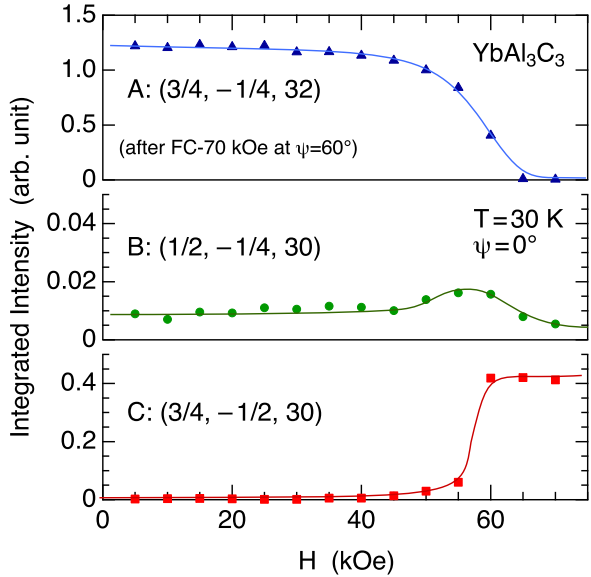}
\end{center}
\caption{(Color online) Magnetic field dependences of the superlattice peak intensities corresponding to the three orthorhombic domains of YbAl$_3$C$_3$ at 30 K after selecting the A-domain by field cooling at 70 kOe at $\psi=60^{\circ}$. 
}
\label{fig:Fig3}
\end{figure}

Switching between different domain states is also possible in the low-temperature orthorhombic phase, which is demonstrated in Fig.~\ref{fig:Fig3}. 
First, the temperature was increased to 90 K above $T_{\text{s}}$ and the sample was rotated to $\psi=60^{\circ}$ ($\mib{H}\parallel [110]_{\text{hex}}$). 
Then, a field of 70 kOe was applied, and the sample was cooled to 10 K at 70 kOe. 
After these procedures, only the A-domain is selected, as expected from the result at $\psi=0^{\circ}$. 
Next, the field was decreased to zero at $\psi=60^{\circ}$, keeping the A-domain state. 
At zero field, the sample was rotated back to $\psi=0^{\circ}$ and the temperature was increased to 30 K, the temperature at which the history-dependent magnetization was measured in Ref.~\citen{Hara12}. 
Then, we measured the intensities corresponding to the three orthorhombic domains with increasing field. 
The results are shown in Fig.~\ref{fig:Fig3}. 

As expected, the sample is in the A-domain state at zero field; the intensities of the B- and C-domains are much smaller than that of the A-domain. 
However, when the field is increased to $\sim 55$ kOe, the intensity of the A-domain abruptly decreases, whereas that of the C-domain increases. This result clearly demonstrates that domain switching from A to C occurred by applying a magnetic field in the orthorhombic phase. 
Note that the intensity of the B-domain slightly increases during the domain switching, although the intensity is kept much smaller than those of the A- and C-domains. 
Thus, we can conclude that the origin of the history-dependent magnetization process reported in Ref.~\citen{Hara12} is the change from a multidomain state to a single-domain state. 

\subsection{DyAl$_3$C$_3$}
\begin{figure}[t]
\begin{center}
\includegraphics[width=8cm]{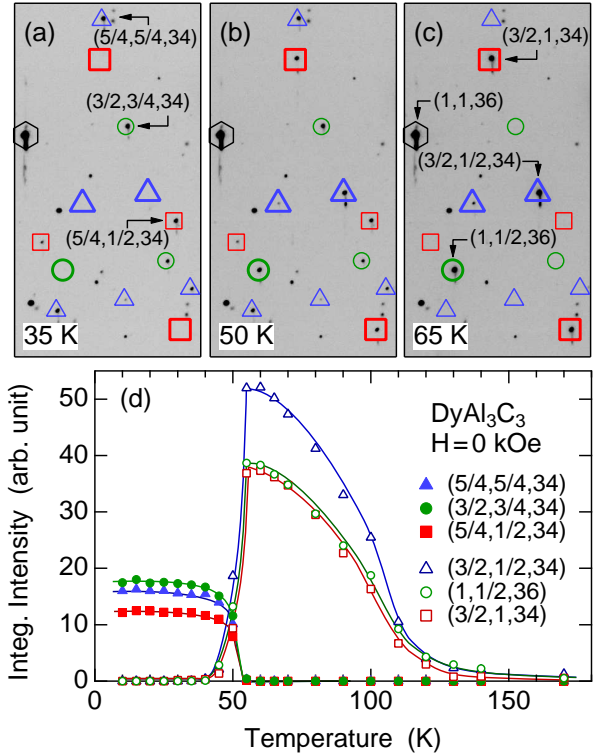}
\end{center}
\caption{(Color online) (a)(b)(c) Aelected area of oscillation photographs of DyAl$_3$C$_3$, demonstrating the disappearance and appearance of the superlattice peaks shown in Figs.~\ref{fig:Fig1}(b) and \ref{fig:Fig1}(c) when crossing the phase boundary at 50 K. 
Small thin marks and large thick marks are peaks appearing below and above 50 K, respectively.  
Hexagons, triangles, circles, and squares represent Bragg spots of fundamental lattice, domain A, domain B, and domain C, respectively. 
(d) Temperature dependences of the peak intensities, which are indexed in (a) and (c). 
}
\label{fig:Fig4}
\end{figure}

Although only one structural transition was suggested at 50 K from the magnetic susceptibility measurement,\cite{Ochiai10} we found two structural transitions in DyAl$_3$C$_3$. 
One is at around 130 K and the other is at 50 K. 
The oscillation photographs, a part of which is shown in Fig.~\ref{fig:Fig4}(a), demonstrate that the low temperature phase below 50 K is characterized by the reciprocal lattice points shown in Fig.~\ref{fig:Fig1}(c), consisting of the same superlattice peaks as those of YbAl$_3$C$_3$. 
Above 50 K, as demonstrated in Figs.~\ref{fig:Fig4}(b) and \ref{fig:Fig4}(c), these peaks disappear and other peaks appear, which can be explained by the reciprocal lattice points shown in Fig.~\ref{fig:Fig1}(b). 
The peaks of all the three domains equally exist at zero field in both phases. 
The temperature dependences of the superlattice peaks, which are indexed in Figs.~\ref{fig:Fig4}(a) and \ref{fig:Fig4}(c), are shown in Fig.~\ref{fig:Fig4}(d). 
Because the intensity exhibits a long tail up to almost 150 K, it is difficult to determine the transition temperature from the intermediate phase to the high-temperature hexagonal phase. 

The successive transitions in DyAl$_3$C$_3$ can be interpreted as the doubling of the orthorhombic unit cell. 
With decreasing temperature, the hexagonal unit cell transforms to the orthorhombic one represented by the dashed rectangles in Fig.~\ref{fig:Fig1}(a). This first transition occurs at approximately 130 K. 
With further decreasing temperature, when crossing the second phase boundary at 50 K, the orthorhombic unit cell is doubled along the $a$-axis, resulting in a larger unit cell represented by  solid rectangles in Fig.~\ref{fig:Fig1}(a).

\begin{figure}[t]
\begin{center}
\includegraphics[width=8cm]{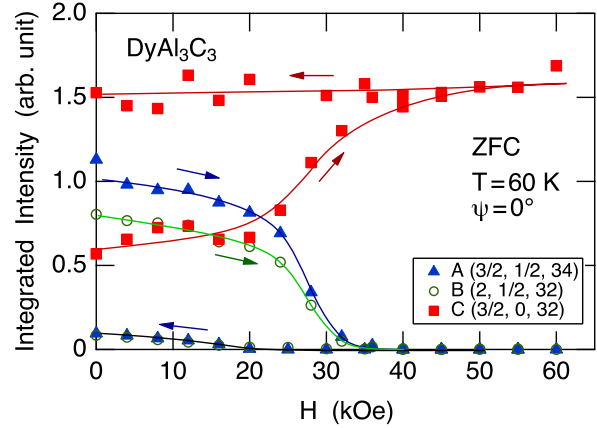}
\end{center}
\caption{(Color online) Magnetic field dependences of the superlattice peak intensities corresponding to the three orthorhombic domains of DyAl$_3$C$_3$ at 60 K in the intermediate phase after zero field cooling. The field is applied along the hexagonal $a$-axis. }
\label{fig:Fig5}
\end{figure}

We next studied the magnetic field effect on orthorhombic domains in DyAl$_3$C$_3$. 
Figure \ref{fig:Fig5} shows the magnetic field dependences of the Bragg peaks corresponding to the three orthorhombic domains at 60 K in the intermediate phase. 
The field was applied at $\psi=0^{\circ}$ ($\mib{H}\parallel \mib{b}_{\text{hex}}$). The sample is initially in the multidomain state because it was cooled at zero field. 
All the peaks from the three domains coexist. 
With increasing field, the intensities corresponding to the A- and B-domains start to decrease at around 25 kOe, where the intensity of the C-domain starts to increase. Finally, at magnetic fields above 50 kOe, only the C-domain remains and the single-domain state is realized. 

When the field is decreased to zero, the intensity of the C-domain remains almost constant, 
whereas those of the A- and B-domains exhibit only a weak increase below 20 kOe. 
This shows that the single-domain state is almost preserved at 60 K even after the magnetic field is removed. 

\begin{figure}[t]
\begin{center}
\includegraphics[width=8cm]{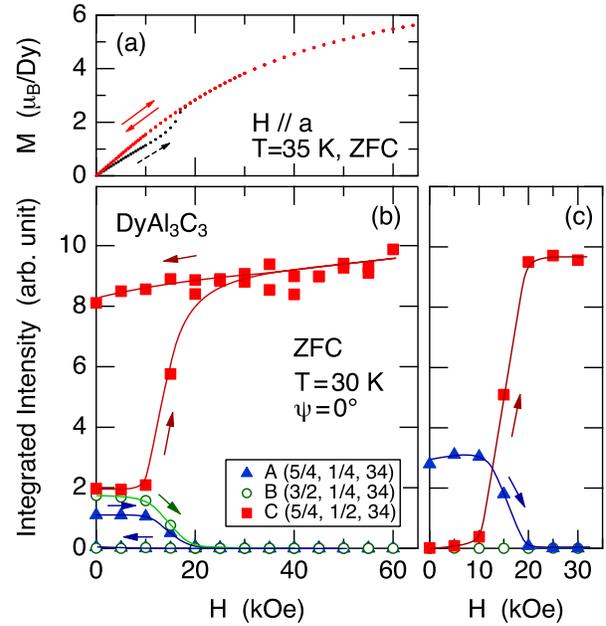}
\end{center}
\caption{(Color online) 
(a) Magnetization of DyAl$_3$C$_3$ at 35 K after zero field cooling. Initially, the $M(H)$ curve follows the process represented by the dashed arrow. After the field is increased to 70 kOe, $M(H)$ follows the curve represented by the solid arrow. 
(b) Magnetic field dependences of the superlattice peak intensities corresponding to the three orthorhombic domains of DyAl$_3$C$_3$ at 30 K in the low-temperature phase after zero field cooling. 
(c) After selecting the A-domain by field cooling at 70 kOe at $\psi=60^{\circ}$. }
\label{fig:Fig6}
\end{figure}

In Fig.~\ref{fig:Fig6}, we show the magnetic field effect on the structural domain state in the low-temperature orthorhombic phase. 
First, Fig.~\ref{fig:Fig6}(a) shows the magnetization at 35 K after ZFC. 
The field direction is along the $a$-axis, which corresponds to $\psi=0^{\circ}$. 
Initially, $M(H)$ follows the process represented by the dashed arrow and exhibits a jump at around 15 kOe. 
After the field is increased once to 70 kOe, $M(H)$ follows the curve represented by the solid arrow in the demagnetization and second magnetization processes. 
Next, Fig.~\ref{fig:Fig6}(b) shows the magnetic field dependences of the superlattice peaks at 30 K after ZFC. 
The data start from the multidomain state, followed by a marked change in intensity at around 15 kOe corresponding to a magnetization jump, and only the C-domain is selected at high fields. 
The magnetization jump at $\sim 15$ kOe, therefore, shows that the multidomain state has changed to the single-domain state.  
Note that the critical field for the domain selection is $\sim 15$ kOe, which is lower than $\sim 25$ kOe at 60 K in the intermediate phase.

Figure~\ref{fig:Fig6}(c) shows the domain switching from A to C at 30 K. 
After the A-domain state was selected by FC(70 kOe) at $\psi=60^{\circ}$, the field was removed, 
the sample was rotated to $\psi=0^{\circ}$, then the field was increased at $\psi=0^{\circ}$. 
In the same way as in YbAl$_3$C$_3$, the structural domain is switched by applying a magnetic field even in the paramagnetic phase. 

\subsection{TmAl$_3$C$_3$}
It is reported that TmAl$_3$C$_3$ exhibits a phase transition at 110 K from the resistivity measurement.\cite{Ochiai10} 
The structural origin of this transition has been confirmed in the present X-ray diffraction experiment. 
The oscillation photographs clearly show that superlattice Bragg peaks appear below 110 K at the reciprocal lattice points shown in Fig.~\ref{fig:Fig1}(b). 
This result shows that the orthorhombic unit cells in the lowtemperature phase are represented by the dashed rectangles in Fig.~\ref{fig:Fig1}(a) and that the three domains equally exist in the ZFC. 
The unit cell is the same as that of DyAl$_3$C$_3$ in the intermediate phase.

\begin{figure}[t]
\begin{center}
\includegraphics[width=8cm]{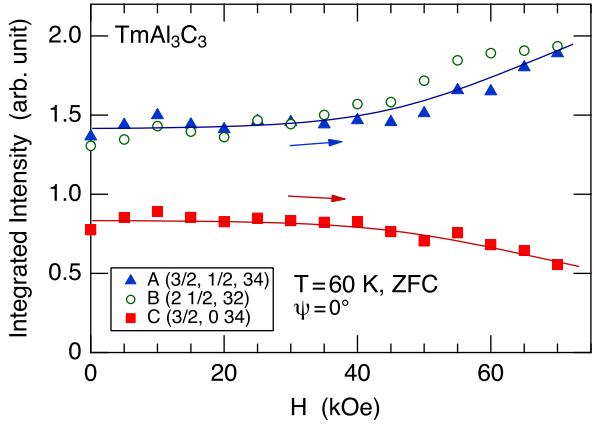}
\end{center}
\caption{(Color online) Magnetic field dependences of the superlattice peak intensities corresponding to the three orthorhombic domains of TmAl$_3$C$_3$ at 60 K in the low-temperature phase. The field is applied along the hexagonal $b$-axis. }
\label{fig:Fig7}
\end{figure}

\begin{figure}[t]
\begin{center}
\includegraphics[width=8cm]{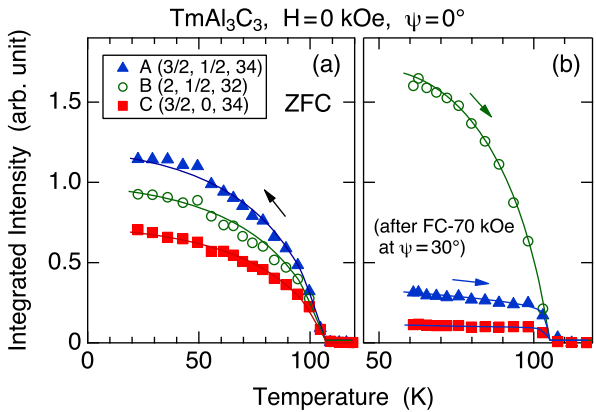}
\end{center}
\caption{(Color online) (a) Temperature dependences of the orthorhombic peak intensities of TmAl$_3$C$_3$ with decreasing temperature at zero field. 
(b) After field cooling at 70 kOe at $\psi=30^{\circ}$. }
\label{fig:Fig8}
\end{figure}

Figure \ref{fig:Fig7} shows the magnetic field dependences of the Bragg peaks corresponding to the three orthorhombic domains  at 60 K in the orthorhombic phase. 
When the field is applied at $\psi=0^{\circ}$, the intensities of the A- and B-domains are enhanced, whereas that of the C-domain is suppressed at high fields. 
This suggests that the orthorhombic $b$-axis prefers to be parallel to the applied field. 
To check this domain selection in TmAl$_3$C$_3$, we investigated the magnetic field effect by applying a field at $\psi=30^{\circ}$. 

As shown in Fig.~\ref{fig:Fig8}(a), all the three orthorhombic domains appear in the ZFC process. 
Next, we increased the temperature to 120 K, rotated the sample to $\psi=30^{\circ}$, applied a magnetic field of 70 kOe, decreased the temperature to 60 K, decreased the field to zero, and rotated the sample back to $\psi=0^{\circ}$. 
Then, the intensities of the three domains were collected with increasing temperature, which is shown in Fig.~\ref{fig:Fig8}(b). 
As expected, the intensity corresponding to the B-domain is largely enhanced whereas those for the A- and C-domains are suppressed. 
Note also that the intensities for the A- and C-domains are still visible in Fig.~\ref{fig:Fig8}(b), which shows that the field of 70 kOe is insufficient to perfectly align the domains in TmAl$_3$C$_3$.

\subsection{LuAl$_3$C$_3$}
LuAl$_3$C$_3$ also exhibits a phase transition at 110 K.\cite{Ochiai10} 
We have confirmed by X-ray diffraction that this is the same type of structural transition as in other RAl$_3$C$_3$ compounds described above. 
The oscillation photographs clearly show that superlattice Bragg peaks appear below 110 K at the reciprocal lattice points shown in Fig.~\ref{fig:Fig1}(b). 
This shows that the orthorhombic unit cells are represented by the dashed rectangles in Fig.~\ref{fig:Fig1}(a), and that the three domains equally exist in the ZFC process. 

\begin{figure}[t]
\begin{center}
\includegraphics[width=8cm]{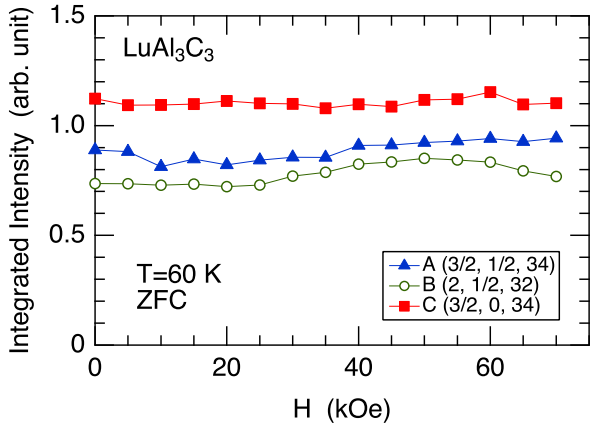}
\end{center}
\caption{(Color online) Magnetic field dependences of the superlattice peak intensities corresponding to the three orthorhombic domains of LuAl$_3$C$_3$ at $\psi=0^{\circ}$.}
\label{fig:Fig9}
\end{figure}

With respect to the magnetic field effect on the orthorhombic domains in the nonmagnetic LuAl$_3$C$_3$, 
as shown in Fig.~\ref{fig:Fig9}, it is difficult to recognise the field dependence of the intensity. 
The intensities are not affected by the external fields, indicating that the multidomain state is preserved up to 70 kOe. 
Then, we conclude that orthorhombic domains are not affected by magnetic fields in LuAl$_3$C$_3$, 
which is different from other magnetic RAl$_3$C$_3$ compounds. 

\section{Discussion} 
\subsection{Orthorhombic structure}
In our previous X-ray diffraction study of YbAl$_3$C$_3$, we concluded that the orthorhombic space group is $Pbca$ (No. 61) from the reflection conditions.\cite{Matsumura08} 
We checked weak reflections on the order of $10^{-6}$ as compared with those of fundamental reflections. 
The reflection conditions obtained in the present study of YbAl$_3$C$_3$ from the oscillation photograph are consistent with those in a previous study, supporting the $Pbca$ space group. 
As pointed out in Ref.~\citen{Khalyavin13}, however, the model structures proposed in Ref.~\citen{Matsumura08} cannot explain the recent neutron powder diffraction data by Khalyavin \textit{et al}. 
This is probably because the atomic positions that we proposed in Ref.~\citen{Matsumura08} are not well optimized. 
At the same time, the model structure proposed in Ref.~\citen{Khalyavin13}, with shifts of Al and C atoms along the $c$-axis only, cannot explain weak reflections, which are clearly observed in the X-ray experiment, such as 22$l$ reflections. 
The atomic displacements in the $c$-plane, allowing all the position parameters of $Pbca$ to be nonzero, are necessary to explain these reflections. 
The determination of the atomic positions remains to be performed in the future. 

The reflection conditions obtained from the oscillation photograph for DyAl$_3$C$_3$ in the low-temperature phase were almost the same as those for YbAl$_3$C$_3$, suggesting that the space group is likely to be $Pbca$. 
With respect to LuAl$_3$C$_3$, TmAl$_3$C$_3$, ErAl$_3$C$_3$, and DyAl$_3$C$_3$ in the intermediate phase, 
the reflection conditions were $h$0$l$ with $l$=even and 00$l$ with $l$=even. We could not obtain reliable data for reflections with $l=0$. 
Then, the candidates for the possible space group are $Pmcm$ (No. 51), $P2cm$ (No. 28), and $Pmc2_1$ (No. 26). 
Here, we refer to an experimental result that, in general, the larger the index $l$, the higher the intensity; it gradually increases with $l$ up to $l$=40, where $\sin \theta / \lambda \sim 1.3$ \AA$^{-1}$. 
This shows that the structure factor is mainly associated with the atomic shifts along the $c$-axis.  
Furthermore, to explain this behavior, it is necessary to include the atomic shifts of R ions. 
Since there is no displacement of R ions in $Pmcm$, and only the displacement of R ions along the $x$-axis is allowed in $P2cm$, it is difficult to reproduce the intensity data with these two space groups. 
Therefore, $Pmc2_1$ (No. 26) is the most likely space group for LuAl$_3$C$_3$, TmAl$_3$C$_3$, ErAl$_3$C$_3$, and DyAl$_3$C$_3$ in the intermediate phase. A detailed structural study is necessary in the future. 

\subsection{Domain motion}
Because we could not identify any change in intensity by the field in nonmagnetic LuAl$_3$C$_3$, it is concluded that the driving force of the domain motion can be ascribed to the magnetic moments of the rare-earth ions. 
Note that the domain selection occurs in the paramagnetic state, where the moment values induced by the field are very small, especially in YbAl$_3$C$_3$. 
Also note that the CEF ground state of TmAl$_3$C$_3$ is a nonmagnetic singlet. 
The domain motion in such a situation might be a nontrivial phenomenon. 

In YbAl$_3$C$_3$, the increase in magnetization ($\Delta M$) from the multidomain state to the single-domain state is 0.01 $\mu_{\text{B}}$ at $B$=6 T and $T$=30 K.\cite{Hara12} 
The difference in Zeeman energy estimated from $\Delta M \cdot B$ is 0.04 K/Yb, which is converted to $3\times 10^{3}$ J/m$^{3}$. 
If we assume this value as an elastic energy, and if we assume a deformation of $\sim 10^{-3}$, a bulk modulus of $\sim$ 10 GPa is obtained, which is not an unusual value. 
Although this is a crude argument, it shows that the magnetic energies of $f$-electrons, which should also include anisotropy energy, can overcome the potential barrier for the orthorhombic domains to move. 
The difference in Zeeman energy estimated as above is of the same order in TmAl$_3$C$_3$ and is 10 times larger in DyAl$_3$C$_3$. 

There are some compounds exhibiting domain switching in the paramagnetic state. 
One is Tb$_2$(MoO$_4$)$_3$, where the crystal axes are interchanged by the field.\cite{Ponomarev94}
This phenomenon can be understood as being caused by a strong magnetoelastic coupling through the quadrupole moment. 
When a magnetic field is applied, the quadrupole moment is induced in the $4f$ orbital. 
Because the quadrupole moment is coupled with the local environment of the lattice, it will induce shifts of the surrounding atoms, which  cooperatively affect the macroscopic lattice parameter, resulting in the interchange between the axes. 
The domain switching in Tb$_2$(MoO$_4$)$_3$ further accompanies a change in the electronic polarization, realizing a magnetoelectric effect. 
A similar case can be seen in PrCu$_2$, where the easy and hard axes of magnetization are interchanged through the reorientation of the quadrupole moment at high fields.\cite{Ahmet96,Settai98} 

In the present case of RAl$_3$C$_3$, it is suggested that the domain selection is caused by the magnetic anisotropy in the $c$-plane, although it is expected to be weak from the result of magnetization.  
In the singlet ground state compound TmAl$_3$C$_3$, the CEF excited states need to be involved through thermal population or Van Vleck terms. 
One possible origin of the anisotropy energy is the magnetoelastic coupling through quadrupole moments induced in magnetic fields. 
However, we have no idea why the selected domain in TmAl$_3$C$_3$ is different from that in YbAl$_3$C$_3$ and DyAl$_3$C$_3$. 
Since the quadrupole moment is related to the sign of the second rank Stevens factor (plus for Tm and Yb; minus for Dy), the anisotropies of Tm and Yb should be the same and that of Dy should be different. 
Another mechanism associated with the spin-orbit interaction of electrons other than $4f$ should also be considered because our preliminary study shows that even GdAl$_3$C$_3$, without a quadrupole moment, possibly exhibits structural domain selection in magnetic fields.\cite{Tosti03}

\subsection{Field-induced magnetic order in YbAl$_3$C$_3$ and magnetic phase diagram}
By neutron powder diffraction in magnetic fields, Khalyavin \textit{et al.} revealed that ferromagnetic and antiferromagnetic components appear at high fields above 6 T at low temperatures.\cite{Khalyavin13}
Strangely, these magnetic signals survive up to $\sim 50$ K, indicating that they are related to the orthorhombic structure below $T_{\text{s}}$. 
The magnetic origin of these signals is inferred from the temperature dependence of the intensity exhibiting a Curie law behavior, which is proportional to the magnetization reported in Ref.~\citen{Hara12}. 
\begin{figure}[t]
\begin{center}
\includegraphics[width=8.5cm]{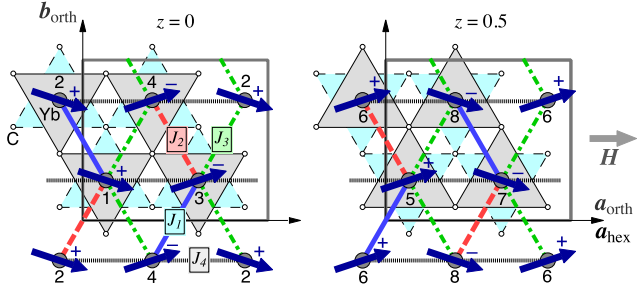}
\end{center}
\caption{(Color online) Schematic model of the field-induced antiferromagnetic structure for $\mib{H}\parallel\mib{a}_{\text{hex}}$ in the low-temperature orthorhombic phase with $Pbca$ space group. 
$+$ and $-$ signs represent the tilt directions of the moments along the $c$-axis. 
The Yb layers on the $c$-planes at $z=0$ and $z=0.5$ are shown with nearest-neighbor C atoms. 
The C atoms connected by the dashed and solid lines are those below and above the Yb layer, respectively. 
In the orthorhombic phase, all the atoms are shifted to the $8c$ sites of the $Pbca$ space group, which results in the production of inequivalent CEFs for the eight Yb sites and induces staggered magnetic moments in magnetic fields. 
The exchange interaction parameters $J_1$, $J_2$, $J_3$, and $J_4$ are shown for reference. 
}
\label{fig:Fig10}
\end{figure}

\begin{table}[t]
\caption{Electric dipole and quadrupole components of the local CEF for the eight Yb atoms in the $Pbca$ space group. 
Although the $e$- and $q$-parameters for the dipole and quadrupole, respectively, are the same for all the Yb atoms, their signs are different. 
The numbers in the Yb column correspond to those in Fig.~\ref{fig:Fig10}. }
\label{tbl:1}
\begin{center}
\begin{tabular}{ccc}
\hline
Yb & $(E_x, E_y, E_z)$ & $(O_{20},O_{22},O_{yz}, O_{zx}, O_{xy})$ \\
\hline
1 & $(-e_x,-e_y,+e_z)$ & $(+q_{20},+q_{22},-q_{yz}, +q_{zx}, +q_{xy})$ \\
2 & $(+e_x,+e_y,-e_z)$ & $(+q_{20},+q_{22},-q_{yz}, +q_{zx}, +q_{xy})$ \\
3 & $(-e_x,+e_y,-e_z)$ & $(+q_{20},+q_{22},-q_{yz}, -q_{zx}, -q_{xy})$ \\
4 & $(+e_x,-e_y,+e_z)$ & $(+q_{20},+q_{22},-q_{yz}, -q_{zx}, -q_{xy})$ \\
5 & $(-e_x,+e_y,+e_z)$ & $(+q_{20},+q_{22},+q_{yz}, +q_{zx}, -q_{xy})$ \\
6 & $(+e_x,-e_y,-e_z)$ & $(+q_{20},+q_{22},+q_{yz}, +q_{zx}, -q_{xy})$ \\
7 & $(-e_x,-e_y,-e_z)$ & $(+q_{20},+q_{22},+q_{yz}, -q_{zx}, +q_{xy})$ \\
8 & $(+e_x,+e_y,+e_z)$ & $(+q_{20},+q_{22},+q_{yz}, -q_{zx}, +q_{xy})$ \\
\hline 
\end{tabular}
\end{center}
\end{table}

The appearance of the antiferromagnetism could be caused by inequivalent CEFs for the eight Yb sites in the $Pbca$ space group. 
When there are no atomic shifts ($P6_3/mmc$), all the Yb ions have the same CEF up to rank 2; there exists only the $O_{20}$ term. 
However, when atomic shifts are introduced according to $Pbca$, all the components of electric dipole and quadrupole fields appear, which are shown in Table \ref{tbl:1}. 
Since the inversion symmetry is lost at the Yb site, an electric dipole field arises. 
If we confine our discussion of CEF to the rank 2 (quadrupolar) term only, four types of CEFs appear in the unit cell; Yb-1 and Yb-2, Yb-3 and Yb-4, Yb-5 and Yb-6, and Yb-7 and Yb-8 have their own CEF.  
In other words, different electric quadrupolar fields appear at the $8c$ sites of Yb.
In such a case, when a magnetic field is applied, a uniform magnetization will necessarily be accompanied by an antiferromagnetic component, as shown in Fig.~\ref{fig:Fig10} by the arrows, because of the inequivalent local CEFs depending on the Yb sites.  
This structure actually gives rise to a nonzero magnetic structure factor for the antiferromagnetic peak reported in Ref.~\citen{Khalyavin13}. 
This is the same situation as that in an antiferroquadrupole ordered state, where the directions of the paramagnetic moments are confined by the ordered quadrupoles.\cite{Link98} 
In this sense, the high-field antiferromagnetic state of YbAl$_3$C$_3$ above 6 T may be interpreted as a normal paramagnetic state of a localized $f$-electron system. 
It will be an interesting subject to study how the nonmagnetic singlet ground state changes to the field-induced antiferromagnetic state. 

The increase in $T_{\text{s}}$ from 80 to 90 K at 30 T, which is reported in Refs.~\citen{Tomisawa08} and \citen{Mito09}, 
could also be explained by considering the magnetoelastic coupling. 
The magnetic field will induce antiferromagnetism with the quadrupole moment, which is compatible with the atomic shifts of Al and C in the $Pbca$ space group. This is expected to help the orthorhombic structure to survive at higher temperatures. 

\section{Conclusion}
We have performed X-ray diffraction experiments in magnetic fields on RAl$_3$C$_3$ for R=Yb, Tm, Dy, and Lu to investigate the hexagonal-to-orthorhombic structural transition and how orthorhombic domains are controlled by external magnetic fields. 
It was clarified that orthorhombic domains in RAl$_3$C$_3$ can be aligned by the fields even in the paramagnetic state if the R ion is magnetic. However, for R=Lu, it was not possible to change the domain populations by the field up to 70 kOe. 
These results show that domain selection and switching are caused by a magnetic anisotropies in the orthorhombic phase of RAl$_3$C$_3$. 

\section*{Acknowledgements}
This work was supported by Grants-in-Aid for Scientific Research (24340087 and 15K05175) from Japan Society for the Promotion of Science (JSPS). 
The synchrotron experiments were performed under the approval of the Photon Factory Program Advisory Committee (No. 2012G-035).

\end{document}